\documentclass{article} 

\usepackage[margin = 2cm]{geometry}
\usepackage{bm}
\usepackage[countmax]{subfloat}
\usepackage{anyfontsize}%
\usepackage{multirow}
\usepackage{footnote}
\usepackage{url}
\usepackage{amsmath}
\usepackage{mathrsfs}
\usepackage{algorithm}%
\usepackage{algorithmicx}%
\usepackage{algpseudocode}%
\usepackage{listings}%
\usepackage{amssymb}
\usepackage{subcaption}
\usepackage{hyperref}
\usepackage{array}
\usepackage{graphicx}
\usepackage{float}
\usepackage{authblk}
\usepackage{natbib}
\newcolumntype{P}[1]{>{\centering\let\newline\\\arraybackslash\hspace{0pt}}p{#1}}

\begin{document}

\title{Precision of Treatment Hierarchy: A Metric for Quantifying Certainty in Treatment Hierarchies from Network Meta-Analysis}
\date{Oct 15, 2024}

\author[1]{Augustine Wigle}

\author[1]{Audrey B\'eliveau}

\author[2]{Georgia Salanti}

\author[3]{Gerta R\"ucker}

\author[3]{Guido Schwarzer}

\author[4]{Dimitris Mavridis}

\author[5,3]{Adriani Nikolakopoulou}

\affil[1]{{Department of Statistics and Actuarial Science}, {University of Waterloo},Canada}

\affil[2]{{Institute of Social and Preventive Medicine}, {University of Bern}, {Switzerland}}

\affil[3]{{Institute of Medical Biometry and Statistics, Faculty of Medicine and Medical Center}, {University of Freiburg},  {Germany}}

\affil[4]{Department of Primary Education, {University of Ioannina}, {Greece}}

\affil[5]{{Laboratory of Hygiene, Social and Preventive Medicine and Medical Statistics, School of Medicine}, {Aristotle University of Thessaloniki},  {Greece}}


\maketitle

\abstract{
Network meta-analysis (NMA) is an extension of pairwise meta-analysis which facilitates the estimation of relative effects for multiple competing treatments. A hierarchy of treatments is a useful output of an NMA. Treatment hierarchies are produced using ranking metrics. Common ranking metrics include the Surface Under the Cumulative RAnking curve (SUCRA) and P-scores, which are the frequentist analogue to SUCRAs. Both metrics consider the size and uncertainty of the estimated treatment effects, with larger values indicating a more preferred treatment. Although SUCRAs and P-scores themselves consider uncertainty, treatment hierarchies produced by these ranking metrics are typically reported without a measure of certainty, which might be misleading to practitioners. We propose a new metric, Precision of Treatment Hierarchy (POTH), which quantifies the certainty in producing a treatment hierarchy from SUCRAs or P-scores. The metric connects three statistical quantities: The variance of the SUCRA values, the variance of the mean rank of each treatment, and the average variance of the distribution of individual ranks for each treatment. POTH provides a single, interpretable value which quantifies the degree of certainty in producing a treatment hierarchy. We show how the metric can be adapted to apply to subsets of treatments in a network, for example, to quantify the certainty in the hierarchy of the top three treatments. We calculate POTH for a database of NMAs to investigate its empirical properties, and we demonstrate its use on three published networks.
}

\newcommand{\sumin}{\sum_{i=1}^n}
\newcommand{\sumjn}{\sum_{j = 1}^n}
\newcommand{\sumint}{\sum_{\substack{i=1 \\ i\neq j}}^n}
\newcommand{\sumkn}{\sum_{k=1}^n}
\newcommand{\sumrn}{\sum_{r = 1}^{n-1}}
\newcommand{\sumrnt}{\sum_{r = 1}^{n-2}}
\newcommand{\sumkr}{\sum_{k = 1}^r}
\newcommand{\sumrm}{\sum_{r = 1}^{m-1}}
\newcommand{\sums}{\sum_{i \in \subsetS}}
\newcommand{\sumkneql}{\sum_{k\neq l}}
\newcommand{\sumjneqi}{\sum_{j \neq i}}
\newcommand{\sumpm}{\sum_{p=1}^m}
\newcommand{\sumlm}{\sum_{l = 1}^m}
\newcommand{\sumrnm}{\sum_{r = 1}^{n/m}}
\newcommand{\sumrnmt}{\sum_{r = n/m+1}^n}
\newcommand{\pik}{p_{ik}}
\newcommand{\pij}{p_{ij}}
\newcommand{\pil}{p_{il}}
\newcommand{\pjk}{p_{jk}}
\newcommand{\pjl}{p_{jl}}
\newcommand{\E}{\text{E}}
\newcommand{\rank}{\text{rank}}
\newcommand{\ranki}{\rank(i)}
\newcommand{\Eranki}{\E(\rank(i))}
\newcommand{\POTH}{\text{POTH}}
\newcommand{\SIR}{\POTH}
\newcommand{\SUCRA}{\text{SUCRA}}
\newcommand{\ssq}{S^2(n)}
\newcommand{\ssqmax}{S^2_{\max}(n)}
\newcommand{\looSUCRAi}{\SUCRA^{-j}(i)}
\newcommand{\sSUCRAi}{\SUCRA^{\subsetS}(i)}
\newcommand{\looSIRi}{\SIR^{-i}}
\newcommand{\looSIRj}{\SIR_{-j}}
\newcommand{\loopik}{p_{ik}^{-j}}
\newcommand{\spik}{p_{ik}^{\subsetS}}
\newcommand{\resj}{r_{\SIR}(j)}
\newcommand{\res}[1]{r_{\SIR}(\text{#1})}
\newcommand{\sssq}[1]{S^2_{#1}(n)}
\newcommand{\Pscore}{\text{P}}
\newcommand{\Pscorei}{\Pscore(i)}
\newcommand{\sPscorei}{\Pscore^{\subsetS}(i)}
\newcommand{\looPscorei}{\Pscore^{-j}(i)}
\newcommand{\thetaij}{\hat{\theta}_{ij}}
\newcommand{\thetaji}{\hat{\theta}_{ji}}
\newcommand{\seij}{SE({\thetaij})}
\newcommand{\sumjnt}{\sum_{\substack{j=1 \\ j\neq i}}^n}
\newcommand{\sumjst}{\sum_{\substack{j \in \subsetS \\ j \neq i}}}
\newcommand{\sSIR}[1]{\SIR_{#1}}
\newcommand{\cSIR}{\text{c}\SIR_{k}}
\newcommand{\subsetS}{\bm{T}}
\newcommand{\subsetSj}{\subsetS^{-j}}
\newcommand{\subsetSbest}{\subsetS^{\text{best }k}}

\section{Introduction}

A hierarchy of treatments is a useful output of a network meta-analysis (NMA) where treatments are ordered from most to least preferred. Ranking metrics are used to produce treatment hierarchies. Common ranking metrics include the Surface Under the Cumulative RAnking curve (SUCRA), the P-score (which is a frequentist version of SUCRA), the expected mean rank, and the probability of having the best outcome value \citep{rucker_ranking_2015, salanti_introducing_2022}. The SUCRAs or P-scores have been shown to be less sensitive to different levels of precision in the effect estimates compared to the probability of having the best value \citep{chiocchia_agreement_2020, davies_degree_2021, salanti_introducing_2022}. 

\begin{figure}[ht]
    \centering
    \includegraphics[width=0.9\linewidth]{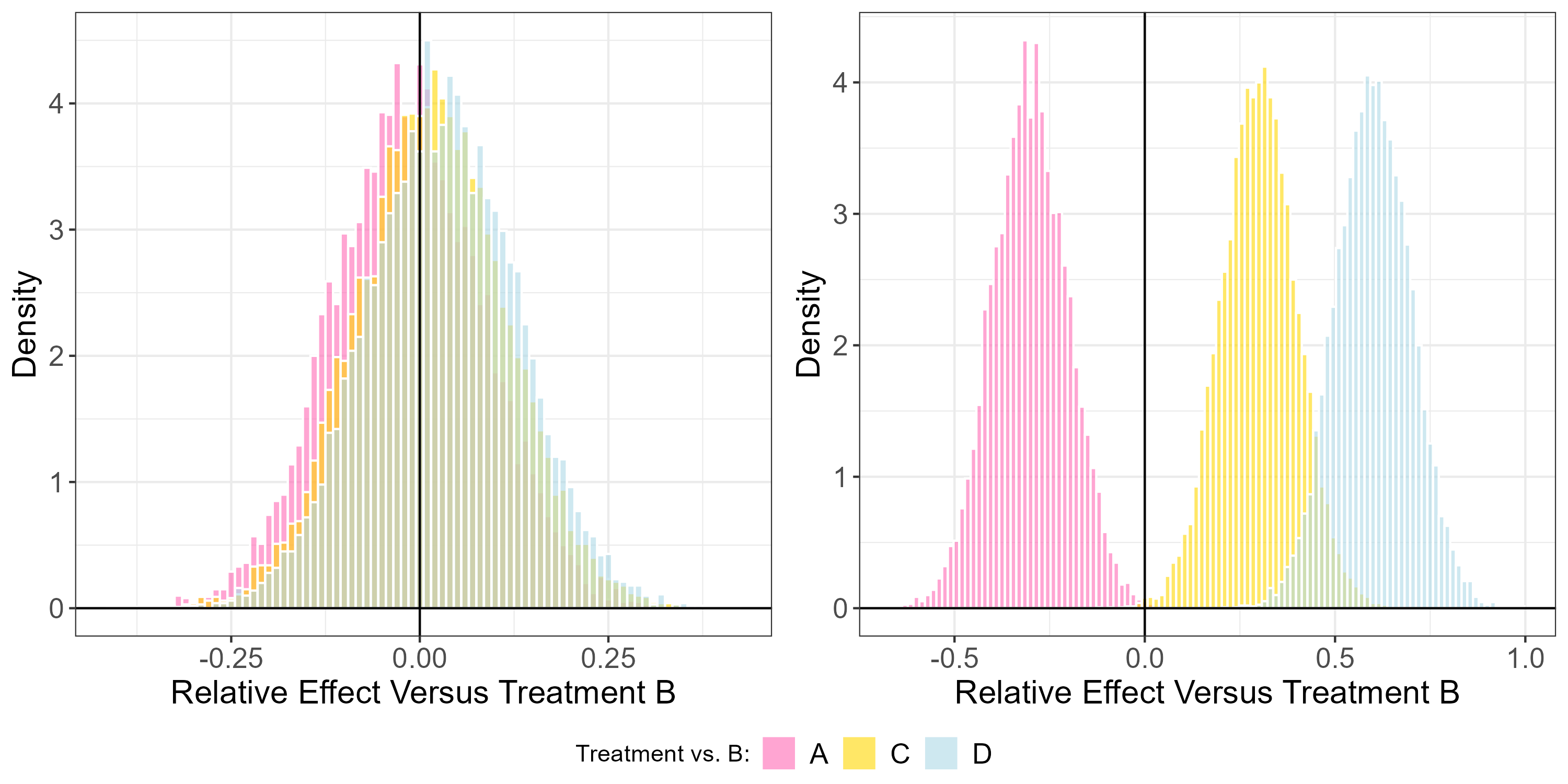}
    \caption{The hypothetical posterior distributions for the relative effect of treatments A, C, and D compared to treatment B that created Sets 1 (left, SUCRA(A) = 0.411, SUCRA(B) = 0.472, SUCRA(C) = 0.530, SUCRA(D) = 0.586) and Set 2 (right, SUCRA(A) = 0.005, SUCRA(B) = 0.334, SUCRA(C) = 0.667, and SUCRA(D) = 0.994).}
    \label{fig:intro-posteriors}
\end{figure}

Treatment hierarchies must be interpreted while considering the underlying uncertainty stemming from the value and precision of the estimated relative effects. For example, consider two hypothetical sets of SUCRA values. Set 1 gives treatments A, B, C, and D SUCRA values of 0.411, 0.472, 0.530, and 0.586, while Set 2 gives SUCRA values of 0.005, 0.334, 0.667, and 0.994. Both sets of SUCRA values would produce the same treatment hierarchy with A the least preferred and D the most preferred. However, the interpretations of the different sets of SUCRAs are very different. The differences in SUCRAs between treatments are much larger in Set 2 than in Set 1. The posterior distribution of the relative treatment effects versus B which produced the SUCRA values for Sets 1 and 2 are shown in Figure \ref{fig:intro-posteriors}. The posterior distributions which produced Set 1 have similar centres and have a lot of overlap with each other and with zero (representing equivalence with treatment B). From a statistical perspective, the posteriors on the left show little evidence of a difference in effects between any of the treatments. On the other hand, the posteriors which produced Set 2 have much less overlap with each other and with zero. In this scenario, we have more confidence that D produces a greater effect than C, and so on. We can therefore interpret Set 1 as producing a hierarchy with low certainty, whereas Set 2 produces a hierarchy with high certainty.

It is necessary to differentiate between the uncertainty in creating a treatment hierarchy and the uncertainty in the ranking metrics used to produce the hierarchy, such as the SUCRA or the P-score. The uncertainty in a given treatment hierarchy is determined by the uncertainty and overlap of estimated relative effects. The uncertainty in SUCRAs or P-scores was discussed by Veroniki et al. \citep{veroniki_is_2018}. The formula for the SUCRA of treatment $i$ is
\begin{equation}
    \SUCRA(i) = \frac{\sumrn\sumkr \pik}{n-1}, \quad i = 1, \dots, n, \label{eq:sucra}
\end{equation}
where $\pik$ is the probability that treatment $i$ takes rank $k$ and $n$ is the number of treatments in the network.
The only uncertainty in $\SUCRA(i)$ comes from the ranking probabilities $\pik$, which are usually assumed to be known. The P-score can be interpreted as an average of p-values, and uncertainty in p-values is rarely considered \citep{veroniki_is_2018}. Moreover, we emphasize that in this work we focus only on the uncertainty in the treatment hierarchy.

Approaches to quantify the uncertainty in producing a treatment hierarchy can broadly be classified as operating at the treatment or hierarchy level. Treatment-level approaches focus on quantifying the uncertainty in the place a certain treatment takes in the hierarchy. Confidence intervals (CIs) for SUCRA values and predictive P-scores are examples of treatment-level approaches. Current approaches to calculating CIs for SUCRA values are restricted to taking discrete values and provide little insight, ranging from zero to one, whereas CIs for predictive P-scores can take continuous values \citep{veroniki_is_2018, wu_using_2021, rosenberger_predictive_2021}. SUCRA and predictive P-score CIs have only been implemented in a Bayesian setting. More recently, it was proposed to use the normalized entropy to quantify uncertainty in each treatment's place in a hierarchy \citep{wu_using_2021}. In a network with $n$ treatments, there will be $n$ SUCRA CIs, predictive P-score CIs, or normalized entropy metrics, each of which may express a different level of certainty. Therefore, treatment-level approaches are not adequate to summarise the overall uncertainty in creating a treatment hierarchy.

Hierarchy-level approaches to quantifying uncertainty have also been suggested. Salanti et al. provided procedures for evaluating the confidence in a hierarchy in the domains defined by the Grading of Recommendations
Assessment, Development and Evaluation (GRADE) Working Group \citep{salanti_evaluating_2014}. These include integrating information from the
contribution matrix and visualizing plots of the ranking probabilities for each treatment (rankograms) \citep{salanti_evaluating_2014}. The methods are not tied to a particular ranking metric. Papakonstantinou et al. suggest approaches for quantifying uncertainty in hierarchies developed from clinically relevant criteria as an alternative to SUCRAs or P-scores \citep{papakonstantinou_answering_2022}. All suggested methods can be subjective or depend on the number of treatments in the network. Finally, since the normalized entropy was proposed as a treatment-level approach, the average normalized entropy among the treatments in the network was used to quantify uncertainty in the treatment hierarchies in an analysis of published NMAs \citep{wu_high_2022}. However, the properties of the average normalized entropy have not yet been explored in depth. No single hierarchy-level approach to quantifying uncertainty has yet been broadly adopted by NMA practitioners.

The aim of this article is to introduce the Precision of Treatment Hierarchy (POTH), a measure of certainty in producing a treatment hierarchy. POTH is derived from SUCRAs or P-scores and operates at the hierarchy-level, summarising the uncertainty in a hierarchy into a single value between zero and one. In Section \ref{sec:methods}, we define scenarios which have maximum and minimum certainty. We introduce the variance of SUCRAs and use this to derive POTH. We also show how POTH is related to the average variance of the treatments' rank probability distributions. We show how POTH can be calculated for subsets of treatments of interest, and discuss two useful applications of POTH for subsets of treatments. In Section \ref{sec:database}, we calculate the POTH in a database of NMAs and investigate associations between the POTH and various network characteristics. In Section \ref{sec:examples}, we provide three worked examples to demonstrate the usefulness and interpretation of POTH. An R package \textbf{poth} and code to reproduce the database analysis and examples are available. We conclude with a discussion of our findings and future work in Section \ref{sec:disc}.



\section{Methodology} \label{sec:methods}

\subsection{Maximum and Minimum Certainty in a Hierarchy} \label{sec:defs}

First, let us clarify the interpretation of ranks and ranking probabilities $\pik$, which are used to calculate SUCRA in Equation \eqref{eq:sucra}. In a Bayesian framework, the $\pik$ are approximated by sampling from the posterior distributions of the relative effects. In a frequentist framework, they are approximated by resampling from the sampling distribution of the relative effects' estimates. A treatment's rank in a given draw from the posterior or sampling distribution is determined by the value of all treatments' relative effects and whether the outcome is beneficial or harmful. In each draw, treatment $i$ will assume one of the $k = 1, 2, \dots, n$ possible ranks. The rank of treatment $i$, $\ranki$, is a random variable. The ranking probability $\pik$ is the probability that treatment $i$ takes the $k^{\text{th}}$ rank in a random draw from the posterior. Together, $\pik$ for $k = 1, \dots, n$ forms the discrete probability distribution for $\ranki$, as shown in Table \ref{tab:rank-dist}.  We treat these probabilities as perfectly known.

\begin{table}[ht]
    \centering
    \caption{The probability mass function for $\ranki$.}
    \label{tab:rank-dist}
    \begin{tabular}{c|cccc}
            Rank & 1 & 2 & \dots & $n$ \\ \hline
            Probability & $p_{i1}$ & $p_{i2}$ & $\dots$ & $p_{in}$
    \end{tabular}
\end{table}

Consider a situation where we are completely certain about the hierarchy of the treatments. In general, this situation is represented by the configuration of ranking probabilities $\pik$ shown on the left in Table \ref{tab:most-certain}. We are completely certain about the hierarchy in this case because the probability of observing the ranking of (1, 2, ..., $n$) in a random draw from the posterior or sampling distribution is one. Note that this situation is unlikely to occur in practice because the order of the relative effects would have to be identical in every sample drawn from the posterior or sampling distribution. In this hypothetical scenario, the SUCRA values are evenly spaced between zero and one by increments of $1/(n-1)$. This is related to the fact that $\SUCRA(i)$ can be expressed in terms of the expected rank of treatment $i$, $\Eranki$ \citep{rucker_ranking_2015}, which is calculated using the probability distribution in Table \ref{tab:rank-dist} as
\begin{equation*}
    \Eranki = \sumkn k\pik,
\end{equation*}
and the SUCRA value of treatment $i$ can be expressed as
\begin{equation} \label{eq:sucra-erank}
    \SUCRA(i) = \frac{n-\Eranki}{n-1}.
\end{equation}
The mean of the SUCRA values in any given network is 0.5 and the mean of the expected ranks in any given network is $(n+1)/2$\citep{rucker_ranking_2015}.


On the other extreme, consider the situation shown on the right in Table \ref{tab:least-certain}. In this scenario, each treatment takes rank $k$ with probability $1/n$ and all SUCRA values are equal to 0.5. This represents a highly uncertain scenario because we have no evidence to suggest ranking any treatment over any others. Other configurations of ranking probabilities are possible which will result in all SUCRA values equal to 0.5. We give illustrative examples in Supplementary Information Section 1. In general, when all SUCRA values are 0.5, any increase in the certainty of the place that one treatment takes in the hierarchy is offset by a decrease in the certainty of another treatment's place in the hierarchy. We therefore define the minimum certainty in a hierarchy as any set of ranking probabilities which results in all SUCRA values equal to 0.5.

In the most certain scenario, the SUCRA values are spread out along the interval $[0, 1]$, whereas in the least certain scenario, the SUCRA values are all equal to their mean. This suggests that deviation of SUCRA values (or mean ranks) from their mean is an indicator of the certainty in a hierarchy.

\begin{table}[ht]
\centering
\caption{Tables of ranking probabilities corresponding to the most certainty in a hierarchy (left) and illustrating the least certainty in a hierarchy (right). The values in the tables correspond to the probability of treatment $i$ taking rank $k$, $\pik$. The bottom rows show the SUCRA values and expected ranks for each treatment.}
\label{tab:most-certain}
\begin{tabular}{c|cccc}
\multirow{2}{*}{Rank ($k$)} & \multicolumn{4}{c}{Treatment ($i$)} \\
 & 1 & 2 & $\dots$ & $n$\\ \hline
{1} & 1 & 0 & $\dots$ & 0\\
{2} & 0 & 1 & $\dots$ & 0 \\
{$\vdots$} & $\vdots$ & $\vdots$ & $\ddots$ & $\vdots$ \\
{$n$} & 0 & 0 & $\dots$ & 1\\ \hline
{SUCRA} & 1 & $(n-2)/(n-1)$ & $\dots$ & 0 \\
$\Eranki$ & $1$ & $2$ & \dots & $n$ \\ 
\end{tabular}
\quad
\label{tab:least-certain}
\begin{tabular}{c|cccc}
\multirow{2}{*}{Rank ($k$)} & \multicolumn{4}{c}{Treatment ($i$)} \\
 & 1 & 2 & $\dots$ & $n$ \\ \hline
{1}          & $1/n$ & $1/n$ & $\dots$ & $1/n$ \\
{2}          & $1/n$ & $1/n$ & $\dots$ & $1/n$ \\
{$\vdots$}   & $\vdots$ & $\vdots$ & $\ddots$ & $\vdots$ \\
{$n$}        & $1/n$ & $1/n$ & $\dots$ & $1/n$ \\ \hline
{SUCRA}& 0.5 & 0.5 & $\dots$ & 0.5 \\
$\Eranki$ & $(n+1)/2$ & $(n+1)/2$ &\dots & $(n+1)/2$\\
\end{tabular}
\end{table}

\subsection{The Variance of SUCRA Values and Mean Ranks}\label{sec:sir}

To investigate the deviation of SUCRAs from their mean, let us consider the variance of the SUCRA values, or equivalently, the scaled variance of the expected ranks. Let
\begin{equation}\label{eq:ssq}
    \ssq = \frac{1}{n} \sumin\left(\SUCRA(i) - 0.5\right)^2
\end{equation}
be the variance of the SUCRA values, which can also be expressed in terms of the variance of the expected ranks using Equation \eqref{eq:sucra-erank} to give
\begin{equation}\label{eq:ssqerank}
   \ssq = \frac{1}{n(n-1)^2}\sumin\left(\Eranki - \frac{n+1}{2}\right)^2.
\end{equation}
Here, we interpret $\ssq$ as a measure of variability of SUCRAs between treatments.
It is clear that when there is minimum certainty in the hierarchy, $\ssq = 0$ because all SUCRA values (or expected ranks) are equal to their mean. In other words, minimum certainty in the hierarchy is identified when the variance of the SUCRA values is zero. It can also be shown that maximum certainty in the treatment hierarchy coincides with maximum variance of the SUCRA values by considering the distribution of each treatment's rank (see Sec. \ref{sec:rank-dist}).

Let $\ssqmax$ be the upper bound of $\ssq$, that is, $\ssqmax$ is the value of $\ssq$ evaluated when the certainty in the hierarchy is maximized and the number of treatments is equal to $n$. Using the fact that in the maximum certainty scenario the SUCRA values form a discrete uniform distribution on $[0, 1/(n-1), 2/(n-1), \dots, 1]$, the maximum variance as a function of $n$ is 
\begin{equation*}
    \ssqmax = \frac{(n+1)}{12(n-1)}.
\end{equation*}
As the number of treatments approaches infinity, this converges to the variance of the uniform distribution from zero to one, which equals $1/12$. 

\subsection{The Precision of Treatment Hierarchy (POTH)}

The upper bound of $\ssq$ is $\ssqmax$, which depends on the number of treatments in the network. This makes $\ssq$ alone difficult to interpret. Therefore, let the Precision of Treatment Hierarchy (POTH) be defined as
\begin{align}
    \SIR(n) = \SIR &= \frac{\ssq}{\ssqmax} \notag \\
    &= \frac{12(n-1)}{(n+1)}\ssq. \label{eq:sir}
\end{align}
Since $\ssq$ is a variance, POTH is non-negative. Thus, we have that 
\begin{equation*}
    0 \leq \POTH = \frac{\ssq}{\ssqmax} \leq 1,
\end{equation*}
with equality to zero when there is minimum certainty in the hierarchy and equality to one when there is maximum certainty in the hierarchy. The POTH is therefore a measure of the degree of certainty in a hierarchy, with higher POTH corresponding to higher certainty. Furthermore, the bounds of POTH do not depend on the number of treatments in the network, facilitating its interpretation.

Note that the key properties of SUCRA values which facilitate the derivation of POTH, namely, a mean of 0.5, are also possessed by frequentist P-scores. P-scores are calculated using the relative effect estimate of $i$ versus $j$, $\thetaij$, and their standard errors $\seij$, assuming a larger response is favourable, according to
\begin{equation}\label{eq:classicpscore}
    \Pscorei = \frac{1}{n-1}\sumjnt \Phi\left(\frac{\thetaij}{\seij}\right),
\end{equation}
where $\Phi(\cdot)$ represents the cumulative distribution function of the standard normal distribution.
If smaller responses are favourable, $\thetaij$ is replaced with $-\thetaij = \thetaji$. Thus, POTH can be calculated using the variance of P-scores by simply replacing $\SUCRA(i)$ with $\Pscorei$ in Equation \eqref{eq:ssq}. Additionally, knowledge of the ranking probabilities $\pik$ is not needed to compute POTH as long as one of the SUCRA values, P-scores, or relative differences and standard errors are available.


\subsection{POTH and the Probability Distribution of Ranks}\label{sec:rank-dist}

We have shown that $\POTH = 0$ uniquely corresponds to the scenario with minimum certainty in the hierarchy. In this section we show that POTH can be calculated from the average variance of the probability distributions of $\ranki$ for $i = 1, \dots, n$ and prove that $\POTH = 1$ uniquely corresponds to the maximum certainty in hierarchy scenario. 

Intuitively, the variance of the distributions of individual ranks are inversely linked to the certainty of a hierarchy. When there is high certainty in the hierarchy, most of the probability will be concentrated at one particular rank for each treatment, leading to distributions of individual ranks with low variances. On the other hand, if the hierarchy is very uncertain, the distributions of individual ranks will be flat with larger variances. Assuming the probabilities $\pik$ are exactly known, the variance of $\ranki$ is given by
\begin{equation*}
    \sumkn \left(k - \Eranki\right)^2\pik.
\end{equation*}
Since we are interested in the uncertainty in the whole hierarchy, we can look at the average of the variances for all observed treatments,
\begin{equation*}
    V(n) = \frac{1}{n}\sumin \sumkn \left(k - \Eranki\right)^2\pik.
\end{equation*}

Under maximum certainty in the hierarchy, $\pik = 1$ for $k = \Eranki$ and $\pik = 0 $ for all $k \neq \Eranki$, for $i = 1, \dots, n$. $V(n)$ is zero under these conditions and it can be shown that this is a unique minimum. Therefore, $V(n)$ is uniquely minimized when certainty in the hierarchy is maximized.

Further, we show that POTH can be expressed in terms of $V(n)$. We can find the relationship between $V(n)$ and $\ssq$ using a weighted sum of squares decomposition as follows:
\begin{equation}\label{eq:Vproof}
    \underbrace{\sumkn \left(k-\frac{n+1}{2}\right)^2}_{\textstyle \frac{n(n+1)(n-1)}{12}} = \underbrace{\sumin\sumkn(k-\Eranki)^2\pik}_{\textstyle nV(n)} + \underbrace{\sumin\sumkn\left(\Eranki - \frac{n+1}{2}\right)^2\pik}_{\textstyle n(n-1)^2\ssq},
\end{equation}
noting that the cross term $2\sumin\sumkn(k-\Eranki)\left(\Eranki-\frac{n+1}{2}\right)\pik$ is equal to zero. Then by dividing Equation \eqref{eq:Vproof} by $n(n+1)(n-1)/12$, we conclude that
\begin{equation}
    \POTH = \frac{12(n-1)}{(n+1)}\ssq = 1-\frac{12}{(n+1)(n-1)}V(n), \label{eq:pothv}
\end{equation}
or in other words, POTH can be derived from $V(n)$ and is equivalent to POTH from the variance of SUCRAs or expected ranks. Since we have shown that $V(n)$ is uniquely minimized when there is maximum certainty in the hierarchy, we also have that $\POTH$ and $\ssq$ are uniquely maximized when there is maximum certainty in the hierarchy. 

\subsection{Deriving POTH for Subsets of Treatments}
Suppose we want to know the certainty in the hierarchy of an arbitrary subset of treatments in a network. For example, clinicians might be interested in the hierarchy of the top five treatments (five treatments with the highest SUCRA values) or the hierarchy of all treatments which are non-surgical. Notice that SUCRA values or P-scores can be re-calculated using subsets of the treatments observed in a network. The interpretation of a SUCRA value or P-score calculated using a subset of the observed treatments is the degree of certainty that a treatment beats other competing treatments, where ``competing treatments" means the subset of treatments considered rather than all treatments in the network. 

To re-calculate SUCRA values, the ranking probabilities $\pik$ must be re-calculated. Let $\subsetS$ represent a subset of all treatments observed in a network. Let $\spik$ be the probability of treatment $i$ taking rank $k$ out of the treatments in $\subsetS$. In a setting where the rank probabilities are calculated from samples of relative effects for treatments $i = 1, . . . , n$, the probabilities $\spik$ can be calculated by ignoring the column of effect estimates for all treatments not in $\subsetS$. Then the SUCRA value for treatment $i$ when the set of competing treatments is defined as $\subsetS$ is given by
\begin{equation}
    \sSUCRAi = \frac{1}{m-1}\sumrm \sumkr \spik,
\end{equation}
where $m$ is the number of treatments in $\subsetS$.

P-scores can be recalculated by simply omitting treatments not of interest from the sum in Equation \eqref{eq:classicpscore}, to give
\begin{equation}
    \sPscorei = \frac{1}{m-1}\sumjst\Phi\left(\frac{\thetaij}{\seij}\right).
\end{equation}
Both $\sSUCRAi$ and $\sPscorei$ maintain a mean of 0.5.

We define the subset POTH of the set $\subsetS$ as
\begin{equation}
    \sSIR{\subsetS} = \frac{12(m-1)}{(m+1)}\sssq{\subsetS},\label{eq:subsetSIR}
\end{equation}
where
\begin{equation*}
    \sssq{\subsetS} = \frac{1}{m}\sums(\sSUCRAi - 0.5)^2.
\end{equation*}
$\sssq{\subsetS}$ could also be calculated with $\sPscorei$.
The interpretation of $\sSIR{\subsetS}$ is the certainty in the treatment hierarchy for the treatments included in $\subsetS$. Subset POTH is always specific to the subset $\subsetS$ and multiple $\sSIR{\subsetS}$ can be calculated for a given network as long as there are more than two treatments in the network. $\sSIR{\subsetS}$ can be higher, lower, or equal to POTH in a given network. In the next two subsections, we describe two useful applications of $\sSIR{\subsetS}$.

\subsubsection{Quantifying Treatment Contributions to POTH} 

\begin{figure}[ht]
    \centering
    \includegraphics[width=0.9\linewidth]{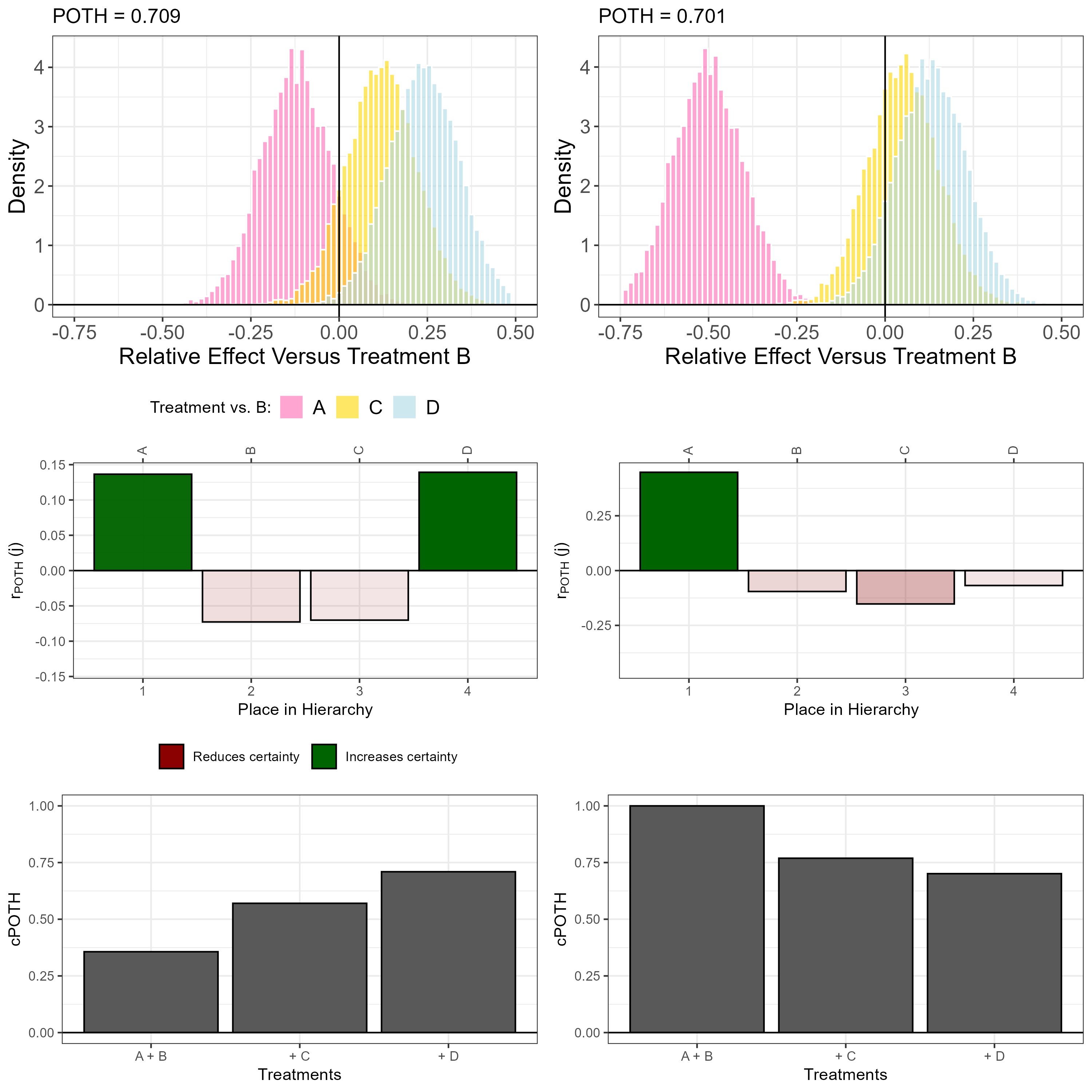}
    \caption{The simulated posterior distributions (top row), POTH residuals $\resj$ (middle row), and $\cSIR$ (bottom row) for two different hypothetical scenarios. Smaller values indicate a better treatment.}
    \label{fig:loo-ex}
\end{figure}

A given value of POTH can be achieved in multiple ways, except for the maximum certainty scenario. For example, a POTH of 0.7 could be observed when the $n-1$ treatments have a similar effect which is considerably different than that of the $n$-th treatment, or, on the other extreme, the $n$ treatments have effects which are moderately different from each other (see example in Figure \ref{fig:loo-ex}). To discern which of the two situations (or any other in-between scenario) is true, it is important to gauge the graphical presentation of the effects and observe the overlap in treatment effects. 
We also introduce here a numerical tool to complement the graphical approach: the leave-one-out POTH residual.

Let $\subsetSj$ be the set of all treatments in the network except treatment $j$, for $j = 1, \dots, n$. Then subset POTH can be calculated to give a set of $n$ leave-one-out POTH values, each one representing the precision in the hierarchy when treatment $j$ is left out. We introduce the POTH residual for treatment $j$,
\begin{equation}
    \resj = \SIR - \sSIR{\subsetSj}. \label{eq:res}
\end{equation}
If 
\begin{equation*}
    \resj < 0
\end{equation*}
then treatment $j$ reduces the certainty in the hierarchy as a whole. Similarly, if
\begin{equation*}
    \resj > 0
\end{equation*}
then treatment $j$ increases the certainty in the hierarchy as a whole. A POTH residual equal to zero means that the certainty in the hierarchy does not change if treatment $j$ is ignored. Under maximum certainty in the hierarchy, the POTH residual for every treatment is zero. Additionally, $\resj$ is bounded between -1 and 1.

\subsubsection{Cumulative POTH} 
The certainty in the best positions in the hierarchy may be of greater interest than that of all treatments. In particular, we may be interested in the POTH of the best $k$ treatments, where $k < n$. We suggest defining $\subsetSbest$ for $k = 2, ..., n$ as the set of  $k$ treatments with the $k$ highest SUCRA or P-score values. Let the cumulative POTH for the best $k$ treatments be
\begin{equation*}
    \cSIR = \sSIR{\subsetSbest}.
\end{equation*}
Note that $\text{c}\SIR_{n} = \SIR$. The cumulative POTH can then be plotted against $k$ to convey how the certainty in the hierarchy changes as $k$ increases to $n$. Significant changes in cPOTH after the addition of a treatment can help identify clusters of similarly performing treatments, as shown in the following hypothetical example. 

\subsubsection{Interpretation of POTH Residuals and Cumulative POTH: A Hypothetical Example}

The POTH residuals and $\cSIR$ in combination with a graphical presentation of the size and precision of estimated effects, such as a plot of the posterior distributions or a forest plot, can aid in interpreting the impact of individual treatments on the certainty in a hierarchy. Consider two networks with four treatments, A, B, C, and D. In one network, the four treatments are all moderately different from each other, while in the other, three treatments B, C, and D perform similarly while A stands out as having a much different effect than the others (for example, A may be a new drug which represents a significant advance over treatment options B, C, and D). We simulate draws from hypothetical posterior distributions under these two scenarios, and calculate POTH, POTH residuals, and cumulative POTH. We assume the outcome is harmful, so smaller responses are desirable. In Figure \ref{fig:loo-ex}, we show the posterior distributions for a situation with moderate separation of all effects (left) and a situation where treatment A is significantly separated from the others (right). Both scenarios have similar POTH (0.709 versus 0.701).

 Below each posterior plot in Figure \ref{fig:loo-ex} we show a graphical representation of the POTH residuals for each treatment. In the situation with similar separation between all treatments, the top and bottom treatments in the hierarchy increase POTH an equivalent amount while the middle two treatments decrease POTH an equivalent amount. The top and bottom treatments increase POTH because they are similar in effects to only one treatment each, while the middle treatments have two neighbouring treatments to which they have similar effects. In the scenario where A is quite different from B, C, and D, the bottom three treatments in the hierarchy reduce the POTH, while the best treatment A, which has a much different effect size than the others, increases the POTH an important amount. 

 In the bottom row of Figure \ref{fig:loo-ex} we present a graphical representation of cumulative POTH for the best $k$ treatments for the two networks.  In the network where all treatments are moderately different, we see a steady increase in $\cSIR$ as more treatments are considered. In the scenario where A is much different from the other treatments, we see the highest $\cSIR$ when only the top 2 treatments, A and B, are considered. This is because the posterior distribution of the relative effect of A vs. B has almost no overlap with zero, giving high confidence in the treatment hierarchy for these two treatments. As more treatments are added, the $\cSIR$ falls since the treatments we are adding have a lot of overlap with each other.

\section{What is the Precision of Treatment Hierarchies in Published Networks? Empirical Evidence from 267 Published Networks}\label{sec:database}

Petropoulou et al. provided a comprehensive database of NMAs published between 1999 and 2015\citep{petropoulou_bibliographic_2017}. From this database, a total of 267 NMAs were identified with available data. POTH was calculated for each network. The R package \textbf{nmadb} was used to access the data \citep{papakonstantinou_nmadb_2019}. Each dataset was re-analysed using frequentist random-effects NMA using the R package \textbf{netmeta} \citep{balduzzi_netmeta_2023}. The effect measure used in the initial publication of each dataset was used in the re-analysis. The REstricted Maximum Likelihood (REML) estimator was used to estimate the between-study heterogeneity parameter $\tau$. P-scores for each treatment were calculated, then POTH was calculated for each network using the R package \textbf{poth}. 

\begin{figure}[ht]
    \centering
        \includegraphics[width = 0.6\linewidth]{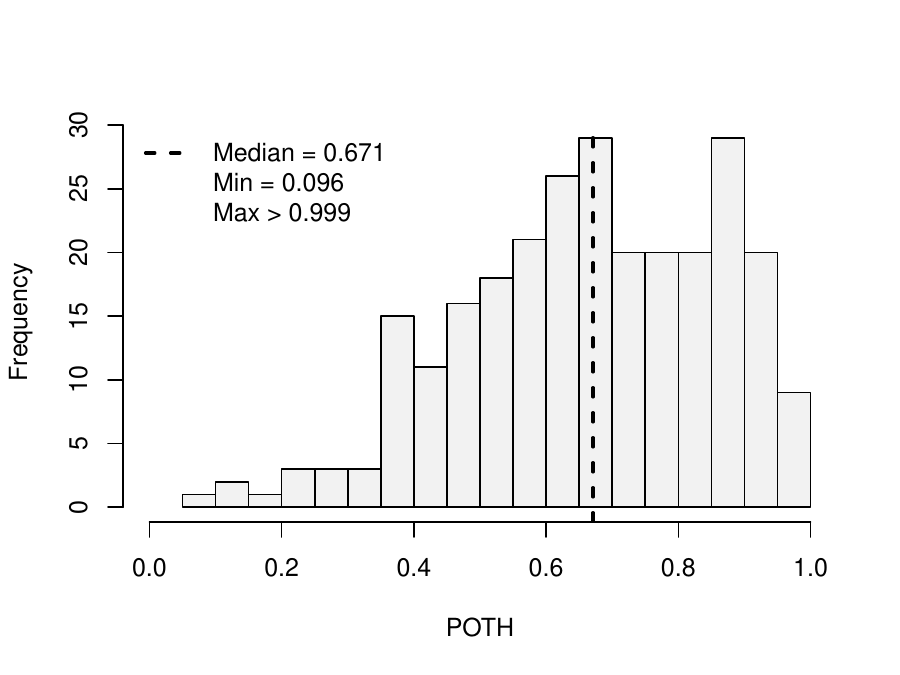}
    \caption{Histogram of POTHs from random-effects NMAs performed on 267 datasets available from \citep{petropoulou_bibliographic_2017, papakonstantinou_nmadb_2019}.}
        \label{fig:sir-dist}
\end{figure}

The distribution of POTH in the networks is shown in Figure \ref{fig:sir-dist}. The minimum POTH is 0.096 and the maximum is over 0.999. Results for the networks producing the minimum and maximum POTH are included in Supplementary Information. The median POTH is 0.671. The inter-quartile range is 0.528 to 0.833.

\begin{figure}[ht]
    \centering
    \includegraphics[width = 0.9\linewidth]{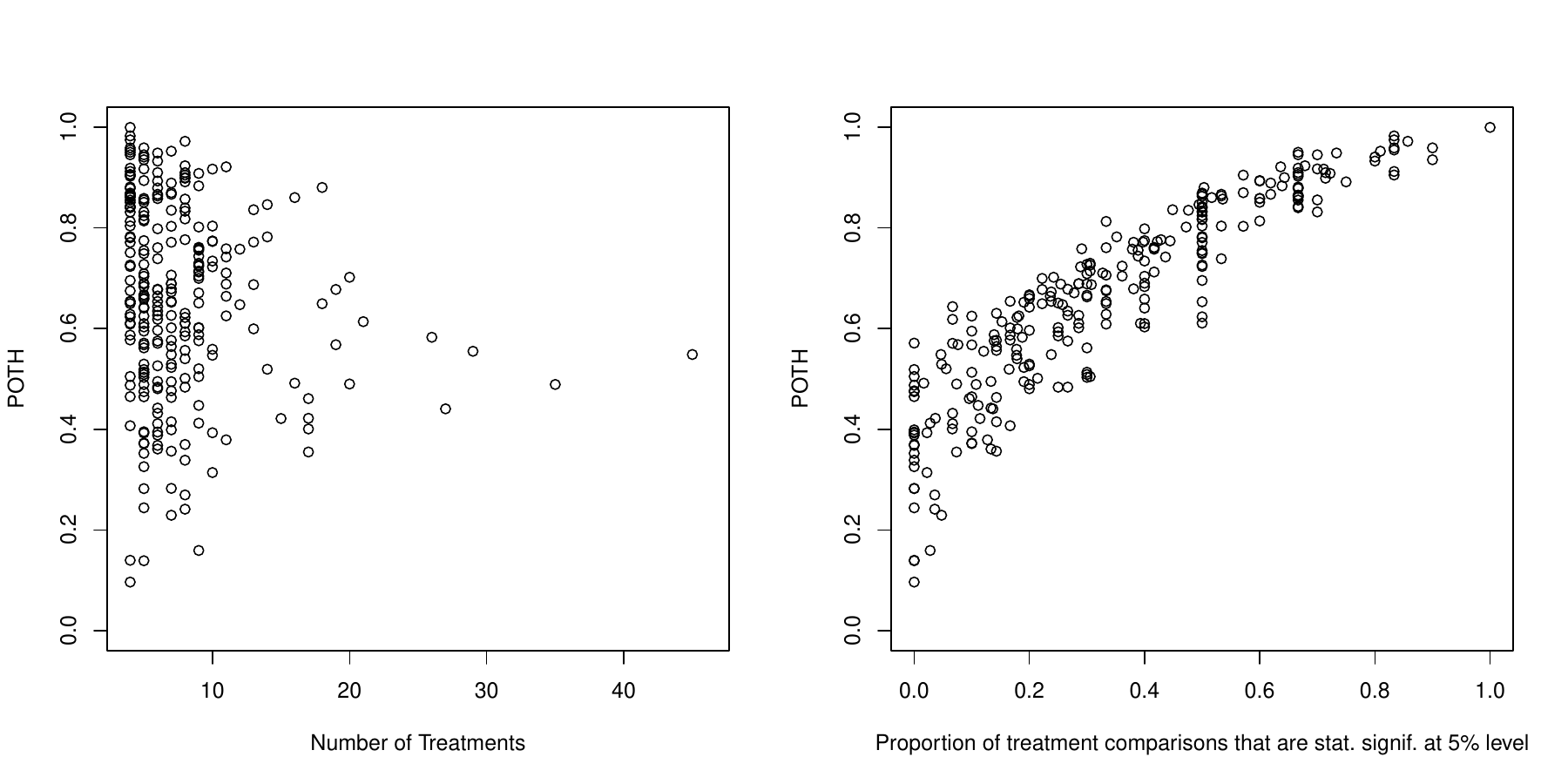}
    \caption{Left: Scatterplot of the POTH versus the number of treatments in the network in the 267 random effects NMAs. Right: Scatterplot of the POTH versus the proportion of treatment comparisons with a p-value less than 0.05 in the 267 random effects NMAs. That is, the x-axis is the proportion of treatment comparisons where there is evidence that the treatments being compared have different effects. }\label{fig:sir-v-ntrt}
\end{figure}

A plot of the POTH versus the number of treatments in the network is shown in the left panel of Figure \ref{fig:sir-v-ntrt}. There are many more NMAs comparing ten or fewer treatments than NMAs comparing more than ten treatments. The Spearman's rank correlation between the number of treatments in the network and the POTH is $-0.193$, indicating a weak relationship. This suggests that we can compare POTH across different networks with varying numbers of treatments. However, there is limited data on the POTH for networks comparing more than 20 treatments, and a better understanding of the behaviour of POTH in networks comparing large numbers of treatments is needed before conclusions about its use in very large networks can be made.

Another interesting relationship is that between the POTH and the number of treatment comparisons which are statistically significant in an NMA. The right panel of Figure \ref{fig:sir-v-ntrt} shows a plot of the POTH versus the proportion of treatment comparisons that are significantly different at the 5\% level. There is a clear relationship between the POTH and the proportion of significantly different treatment comparisons, with POTH values close to one tending to have a proportion of significant comparisons close to one as well. The converse is true to some extent, however there is more variation. For example, among NMAs where there were zero significantly different treatment comparisons, the range of POTH values is from 0.096 to 0.571. Still, this association can aid in the interpretation of POTH values. For a network with POTH close to one, we expect almost all treatment comparisons to be significant. For a network with POTH less than or equal to 0.5, the proportion of significant treatment comparisons is expected to be much lower - the largest proportion of significant treatment comparisons seen in networks with POTH less than 0.50 was 0.267. 

\begin{figure}[ht]
    \centering
    \includegraphics[width=\linewidth]{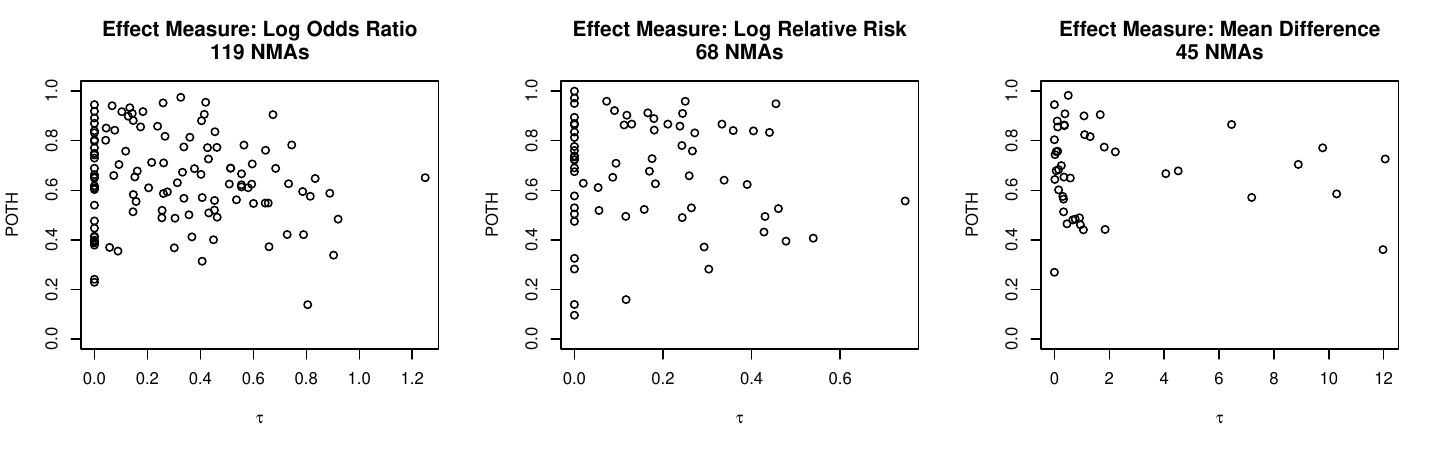}
    \caption{Plots of POTH versus the estimated heterogeneity parameter $\tau$ for the three most common effect measures in \textbf{nmadb}: log odds ratio, log relative risk, and mean difference. There is an outlying point that was omitted from the mean difference plot which had an estimated $\tau$ of 68.06 and POTH equal to 0.597.}
    \label{fig:sir-v-tau}
\end{figure}

Additionally, we investigate the relationship between POTH and heterogeneity in the network. Since the interpretation of the heterogeneity parameter $\tau$ depends on the effect measure, we investigate the NMAs based on the effect measure used in the analysis. The three most common effect measures in the database were the log odds ratio (119 NMAs), log relative risk (68 NMAs), and mean difference (45 NMAs). There were 14 or fewer NMAs using each of the remaining effect measures. The POTH versus the estimate of $\tau$ is shown for the top three effect measures in Figure \ref{fig:sir-v-tau}. The Pearson's correlation coefficient between POTH and $\tau$ for each effect measure are -0.143, -0.074, and -0.110 for log-odds ratio, log relative risk, and mean difference respectively. This indicates a weak negative correlation between POTH and the heterogeneity of a network. 

\section{Three Worked Examples} \label{sec:examples}
Here we present three published networks to demonstrate the use and interpretation of POTH, POTH residuals and cumulative POTH. In each example, a random-effects NMA model was fit to the data using the REML method to estimate heterogeneity. The POTH was calculated using the R package \textbf{poth}.

\subsection{Antifungal Treatments to Prevent Mortality for Solid Organ Transplant Recipients}
A meta-analysis of treatments for preventing fungal infection in solid organ transplant recipients was conducted by Playford et al. \citep{playford_antifungal_2004}. They report data from ten studies. There are five treatments to compare; four anti-fungal treatments and a control. The data is available in \textbf{nmadb} using record ID 501370. \citep{papakonstantinou_nmadb_2019}. The outcome of interest is mortality.

\begin{figure}[ht]
    \centering
    \begin{subfigure}{0.9\linewidth}
        \includegraphics[width = \linewidth]{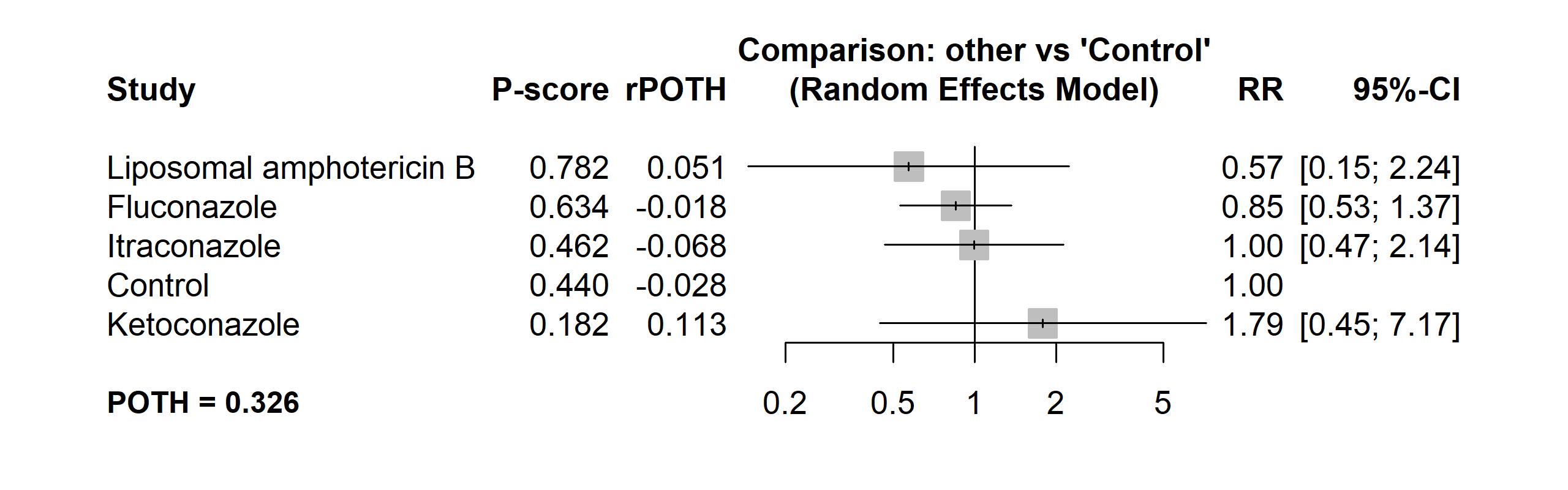}
    \end{subfigure}
    \begin{subfigure}{0.9\linewidth}
        \includegraphics[width = \linewidth]{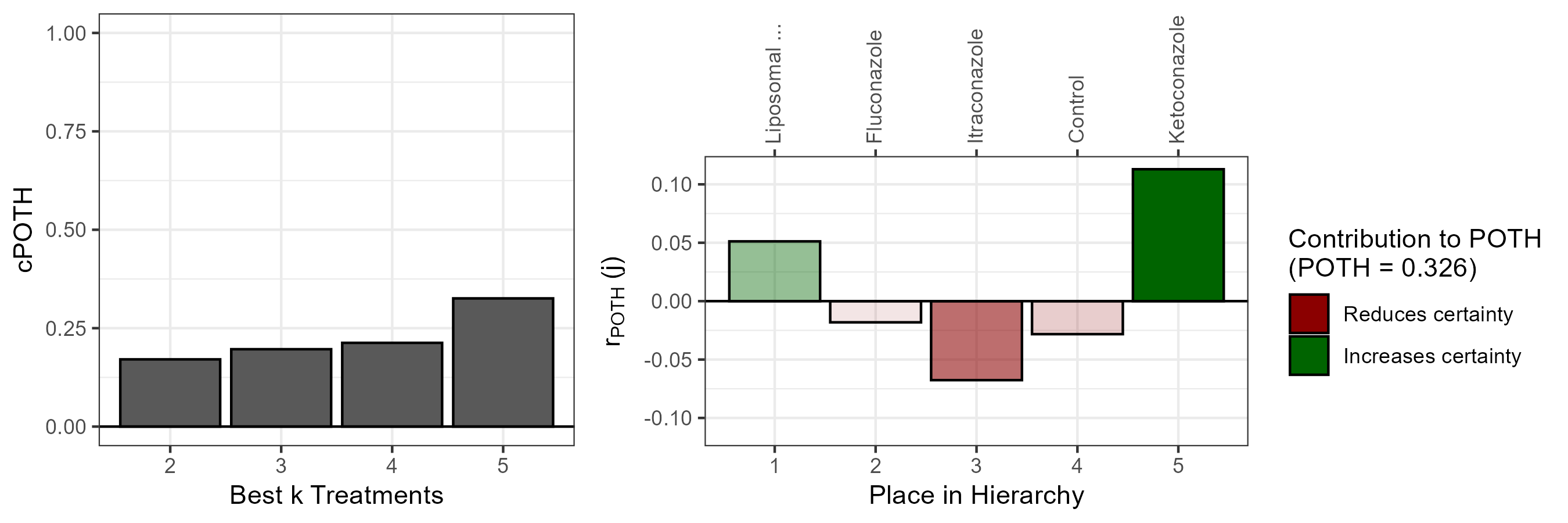}
    \end{subfigure}
    \caption{Above: Estimated confidence intervals (CI) for the relative risk of mortality for anti-fungal treatments versus control in the organ transplant network. P-scores and residual POTH are shown to the left of CIs. RRs less than one favour the intervention and RRs greater than one favour control. The treatments are organized from largest (highest ranking in hierarchy) to smallest (lowest ranking in hierarchy) P-scores. A positive residual indicates that the treatment contributes to the certainty in the hierarchy, while a negative residual indicates that the treatment reduces the certainty in the hierarchy.\\
    Below, left: Cumulative POTH for the best two to $n$ treatments in the organ transplant network. \\
    Below, right: POTH residuals $\resj$ for all treatments in the network.}
    \label{fig:forest-transp}
\end{figure}

The 95\% CIs, P-scores, and POTH residuals from the random-effects NMA for the transplant data are shown in Figure \ref{fig:forest-transp}. The POTH calculated using the P-scores is 0.326, reflecting low certainty in the treatment hierarchy. For reference, in the database analysis in Section \ref{sec:database}, 95.5\% of networks had a POTH greater than 0.326. Looking at the CIs, it is clear that there is a lot of overlap in the sampling distributions of the treatment effects compared to control, so the low POTH is not surprising. Turning to the POTH residuals, we note that Ketoconazole, the lowest-ranked treatment in the hierarchy, has the largest positive residual equal to 0.113. Despite its wide CI, the estimated relative effect of Ketoconazole is noticeably different from those of the other treatments, so overall it adds certainty to the hierarchy. The cumulative POTH for the best $k$ treatments, $k = 2, \dots, 5$ is illustrated in the bottom panel of Figure \ref{fig:forest-transp}. There is a steady increase in POTH as more treatments are considered, with the largest jump occurring when $k$ increases from $4$ to $5$, corresponding to the addition of Ketoconazole to the set of competing treatments.

Suppose we wish to produce a hierarchy of all non-control treatments in addition to the hierarchy of all treatments. We use $\sSIR{\subsetS}$ to quantify the certainty in the hierarchy of all non-control treatments by defining $\subsetS$ = \{Liposomal amphotericin B, Fluconazole, Itraconazole, and Ketoconazole\}. The $\sSIR{\subsetS}$ for the non-control treatments is 0.354. In fact, this corresponds to the POTH residual for control. Removing control from the set of treatments leads to an increase in POTH of 0.028. There is slightly more certainty in the hierarchy of non-control treatments compared to the hierarchy of all treatments, but the certainty is still low.

\subsection{Efficacy of Pharmacological Treatments for Persistent Depressive Disorder}

Kriston et al. conducted an NMA assessing the efficacy and acceptability of treatments for persistent depressive disorder \citep{kriston_efficacy_2014}. Data from 45 studies comparing 29 treatments (28 pharmacological treatments, and placebo) were accessed in \textbf{nmadb} from record ID 480666 \citep{papakonstantinou_nmadb_2019}. The binary outcome of interest is an improvement of 50\% or more on a symptom severity scale, so larger values are desirable. 

\begin{figure}[ht]
    \centering
    \includegraphics[width=0.9\linewidth]{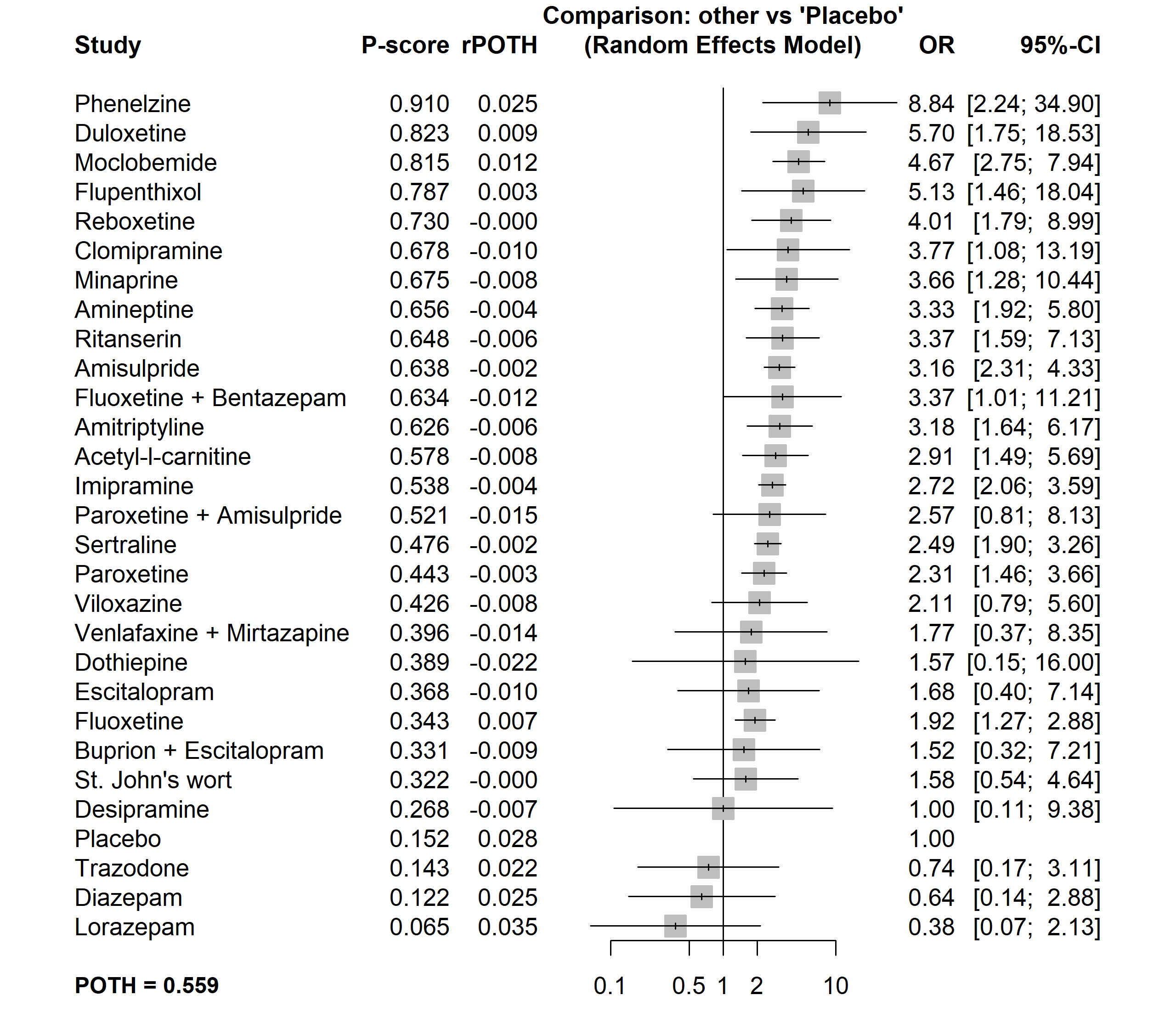}
    \caption{Estimated confidence intervals (CI) for the odds ratio of 50\% improvement of symptom severity for pharmacological interventions versus placebo in the depression network. P-scores and residual POTH are shown to the left of CIs. ORs greater than one favour the intervention and ORs less than one favour placebo. The treatments are organized from largest (highest ranking in hierarchy) to smallest (lowest ranking in hierarchy) P-scores. A positive residual indicates that the treatment contributes to the certainty in the hierarchy, while a negative residual indicates that the treatment reduces the certainty in the hierarchy.}
    \label{fig:forest-depression}
\end{figure}

A forest plot in Figure \ref{fig:forest-depression} shows the results of the random-effects NMA. The treatments are organized by decreasing P-score. The POTH is calculated as 0.559. This represents moderate certainty in the ranking. In our analysis of the database, 71.5\% of networks had a greater POTH than 0.559. 

\begin{figure}[ht]
    \centering
    \includegraphics[width=0.9\linewidth]{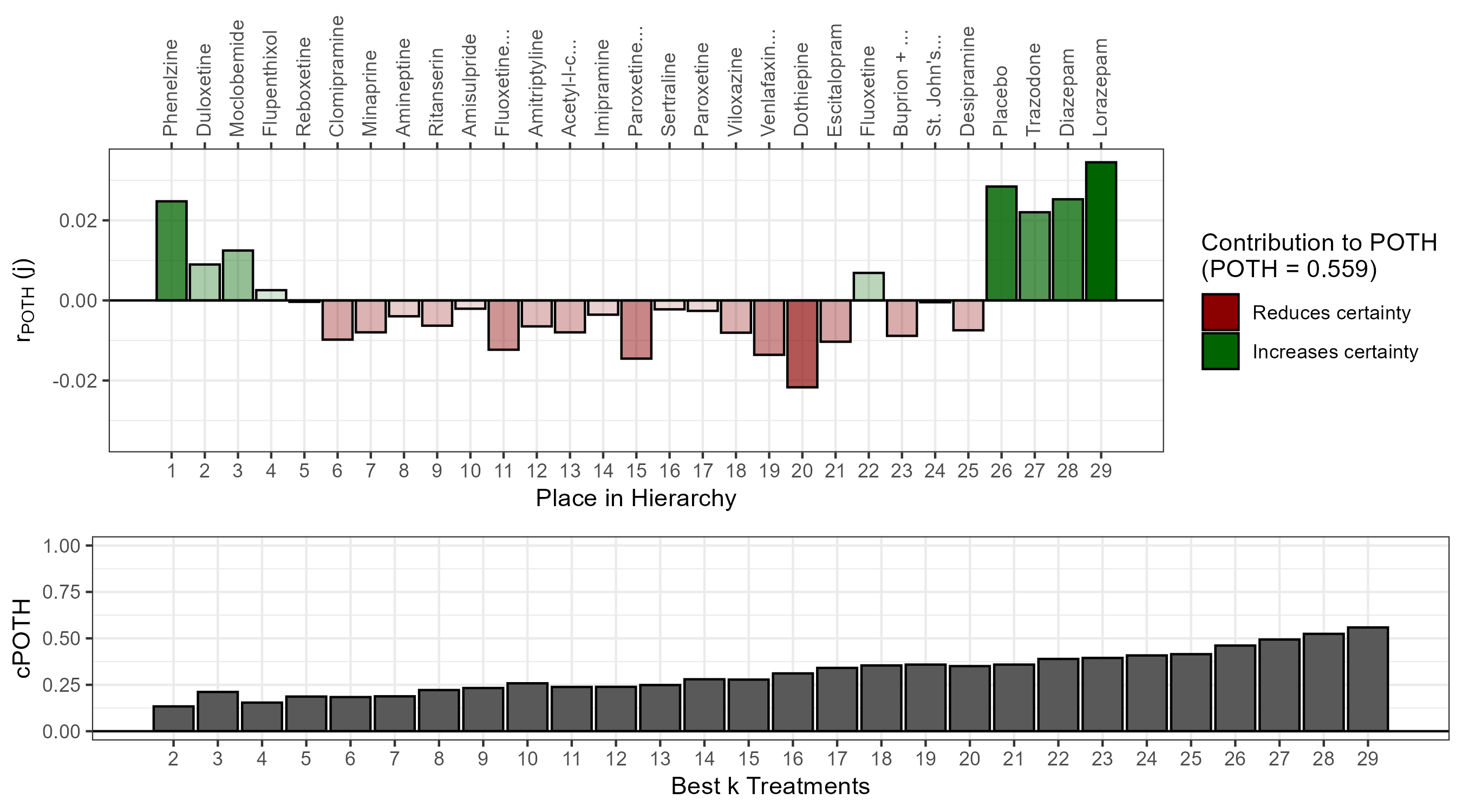}
    \caption{Left: Cumulative POTH for the best two to $n$ treatments in the depression network. \\
    Right: POTH residuals $\resj$ for all treatments in the network.}
    \label{fig:cum-depression}
\end{figure}

The POTH residuals for all 29 treatments and the cumulative POTH for the top $k$ treatments in the depression network for $k = 2, \dots, 29$ is illustrated in Figure \ref{fig:cum-depression}. Based on the $\resj$, the treatments contributing the most to POTH are the four worst treatments in the hierarchy, and the top treatment in the hierarchy, phenelzine. Dothiepine is the treatment with the lowest POTH residual equal to -0.022, indicating that it reduces the certainty in the hierarchy the most out of all treatments in the network. This is no surprise as it has a very wide CI and similar point estimates to neighbouring treatments in the hierarchy. As more treatments are considered, the POTH rises relatively steadily. This is reflective of the lack of clustering of estimated treatment effects as seen in the forest plot. The cumulative POTH is the lowest when considering only the top 2 treatments, phenelzine and duloxetine.

\subsection{Safety of Immune Checkpoint Inhibitors as Cancer Treatments}
A NMA of Immune Checkpoint Inhibitors (ICIs) was conducted by Xu et al.  to assess the safety of ICI drugs as cancer treatment \citep{xu_comparative_2018}. Data were obtained from Rosenberger et al., who re-analysed the data\citep{rosenberger_predictive_2021}. There are seven treatments and 23 studies. The outcome of interest is the number of treatment-related adverse events, so smaller values indicate a safer treatment.

\begin{figure}[ht]
    \centering
    \begin{subfigure}{0.9\linewidth}
        \includegraphics[width=\linewidth]{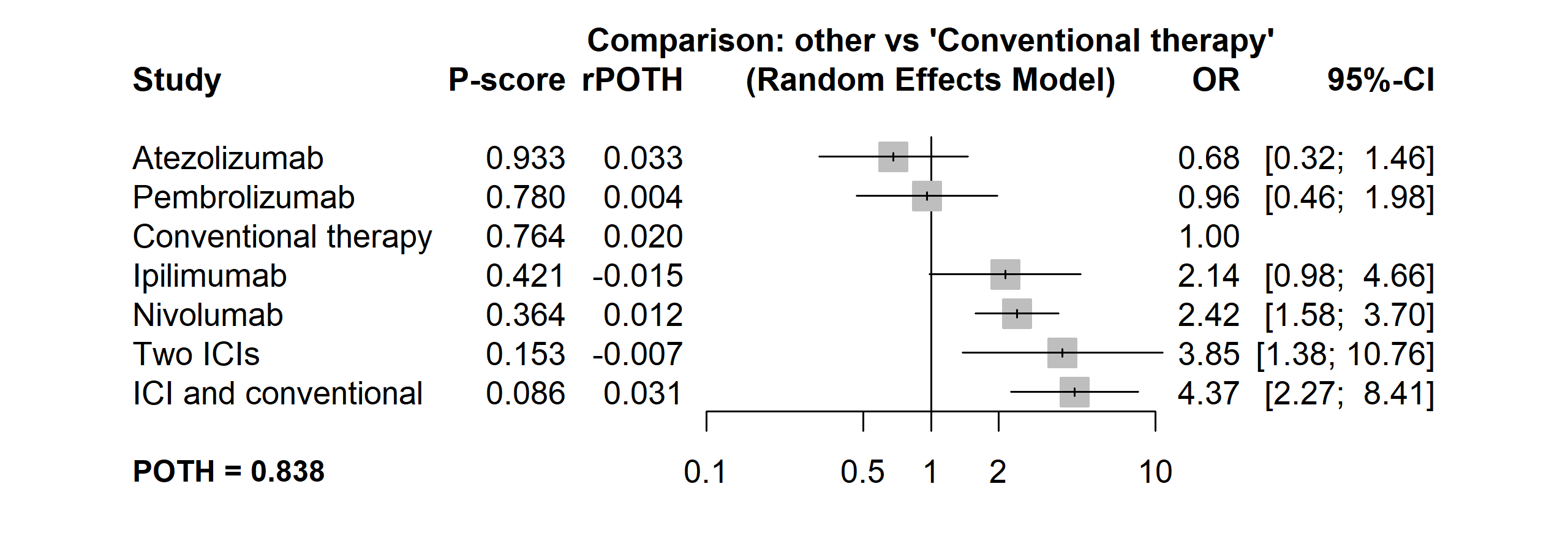}
    \end{subfigure}
    \begin{subfigure}{0.9\linewidth}
        \includegraphics[width=\linewidth]{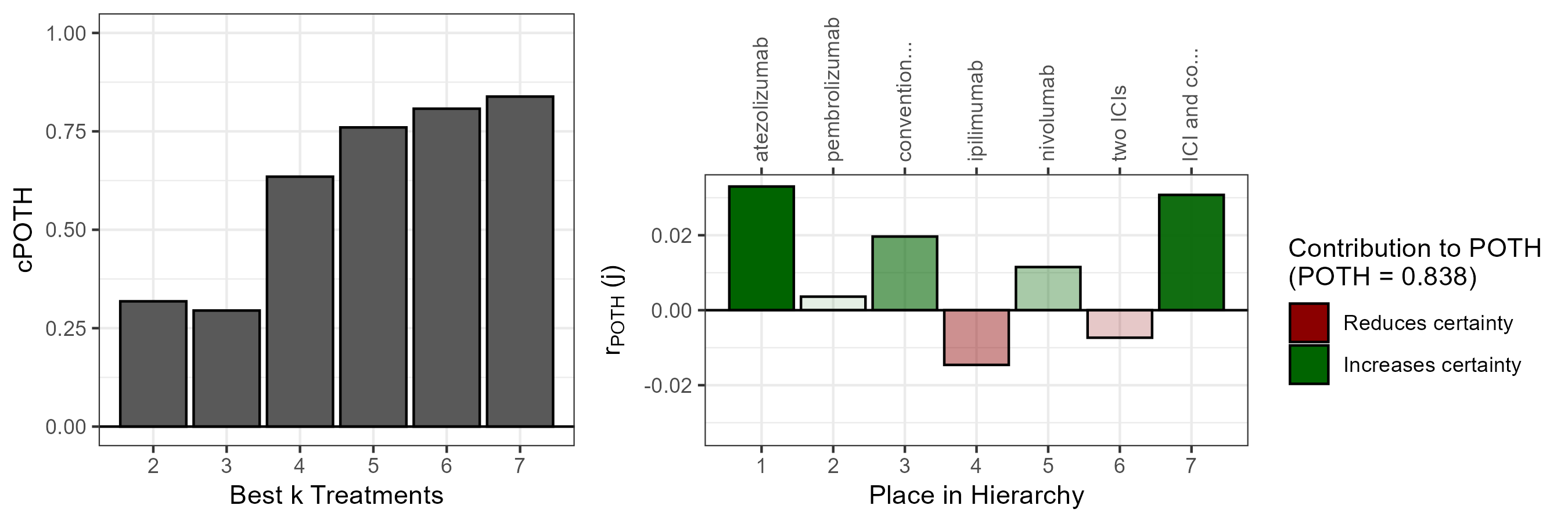}
    \end{subfigure}
    \caption{Above: Estimated confidence intervals (CI) for the odds ratio of all-grade treatment-related adverse events for ICI versus conventional therapy in the cancer network. P-scores and residual POTH are shown to the left of CIs. ORs less than one favour the intervention and ORs greater than one favour conventional therapy. The treatments are organized from largest (highest ranking in hierarchy) to smallest (lowest ranking in hierarchy) P-scores. A positive residual indicates that the treatment contributes to the certainty in the hierarchy, while a negative residual indicates that the treatment reduces the certainty in the hierarchy. \\
    Below, left: Cumulative POTH for the best two to $n$ treatments in the ICI network. \\
    Below, right: POTH residuals $\resj$ for all treatments in the network.}
    \label{fig:forest-ici}
\end{figure}

The estimated treatment effects are shown in the forest plot in Figure \ref{fig:forest-ici}. Atezolizumab is ranked as the safest in the hierarchy and ICI and conventional is ranked as the least safe. The POTH is equal to 0.838, indicating moderate to high certainty in the treatment hierarchy. Compared to the database of networks, 24.0\% of networks had POTH greater than 0.838. 

The POTH residuals are shown in the column beside the P-scores and in the bottom right panel. The treatments contributing the most to certainty in the hierarchy are atezolizumab, ICI and conventional, and conventional therapy. Ipilimumab and two ICIs reduce the certainty in the hierarchy. Ipilimumab has a similar point estimate to nivolumab, and two ICIs has the widest CI in the network.

The cumulative POTH for the top $k$ treatments is shown in the bottom panel of Figure \ref{fig:forest-ici}. The POTH is the lowest when the top three treatments are considered. There is a noticeable increase in the cPOTH when going from the top three to the top four treatments. The certainty then increases steadily as more treatments are added. This illustrates that there is a cluster of treatments with similar safety profiles occupying the top three spots in the hierarchy made up of atezolizumab, pembrolizumab, and conventional therapy. Indeed, we can see the similarity in the CIs of atezolizumab and pembrolizumab versus conventional therapy in the forest plot in Figure \ref{fig:forest-ici}.

Finally, consider creating a hierarchy for only the treatments which are single ICIs. That is, let $\subsetS = $\{atezolizumab, pembrolizumab, ipilimumab, nivolumab\}. The resulting $\sSIR{\subsetS}$ is 0.777. This is lower than the POTH for all treatments in the network (0.838), indicating that there is more certainty in the hierarchy considering all treatments in the network than in the hierarchy of only the single ICIs. In Figure \ref{fig:forest-ici}, we can see that the single ICIs roughly form two clusters (atezolizumab/pembrolizumab and ipilimumab/nivolumab). In Supplementary Information section 2, we derive the POTH for a hypothetical example where the treatments are divided into an arbitrary number of equally sized clusters where all treatments in the same cluster have the same performance. This example is reminiscent of the case when there are two clusters, where we found that the POTH is bounded below by 0.75.


\section{Discussion}\label{sec:disc}

In this paper, we introduce POTH, a metric which quantifies the certainty in producing a treatment hierarchy. We define maximum and minimum certainty scenarios and show that POTH is uniquely maximized in the case of maximum certainty and minimized in the case of minimum certainty. We propose $\sSIR{\subsetS}$ as a method to measure certainty in the hierarchy of a subset of treatments. The POTH answers the question ``how certain is the presented hierarchy of all treatments?" while $\sSIR{\subsetS}$ answers the question ``how certain is the hierarchy of a subset of treatments of interest?". We also introduce two useful applications of $\sSIR{\subsetS}$: POTH residuals using a leave-one-out approach to measure the contribution of each treatment to POTH, and cumulative POTH to identify the certainty of the hierarchy for the top $k$ treatments as $k$ increases. The methods are easily applied in Bayesian and frequentist frameworks.


POTH is a quantitative approach, which sets it apart from some previously suggested approaches to quantifying uncertainty in treatment hierarchies \citep{papakonstantinou_answering_2022, salanti_evaluating_2014}. The average normalized entropy is also a quantitative hierarchy-level approach. An advantage of POTH over the average normalized entropy is that POTH can be calculated directly from SUCRAs or P-scores rather than relying on the individual ranking probabilities $\pik$. We have also explored the properties of POTH via the database analysis in Section \ref{sec:database}. 

We performed a preliminary simulation study to investigate the distribution of POTH that can be expected in networks where no true treatment differences exist. The results are available in Supplementary Information. However, a more extensive simulation study which investigates the impact of different magnitudes of treatment differences, precision in the data, number of treatments, network geometries, and heterogeneity is an important area of future work. 


 Although POTH was proposed specifically to summarise the certainty in treatment hierarchies derived using SUCRAs or P-scores, it can also be viewed simply as a way to summarise all the ranking probabilities in a given network. It is therefore an indicator of the certainty in any treatment hierarchy derived using a ranking metric related to the ranking probabilities.
A strength of POTH is that it can be applied to any extensions of SUCRA or P-values, as long as their range is zero to one and their mean is 0.5. For example, a version of POTH could be calculated using Bayesian predictive P-scores to quantify the certainty in the hierarchy of treatments when applied in a new study \citep{rosenberger_predictive_2021}. 

It was previously highlighted that the quality of evidence in NMA should be evaluated separately for the treatment effect estimates and the hierarchy of treatments \citep{salanti_evaluating_2014}. Formal guidance for assessing the certainty in NMA estimates has been provided, notably GRADE \citep{guyatt_grade_2008, guyatt_grade_2011} and Confidence in Network Meta-Analysis (CINeMA)\citep{nikolakopoulou_cinema_2020}. However, no similar guidance for the hierarchies resulting from an NMA have been created. POTH and POTH for subsets of treatments could play a role in the future development of guidelines for evaluating the certainty in treatment hierarchies from NMA.

\section{Acknowledgements}

AW acknowledges the support of the Natural Sciences and Engineering Research Council of Canada (NSERC), Grant numbers CGS-D 569445-2022, CGS-MSFSS 588256-2023. AB was supported by NSERC Discovery Grant RGPIN-2019-04404. DM was supported by the framework of H.F.R.I call ``Basic Research Financing (Horizontal support of all Sciences)" under the National Recovery and Resilience Plan ``Greece 2.0" funded by the European Union - NextGenerationEU (H.F.R.I. Project Number: 015467). AN was supported by the Deutsche Forschungsgemeinschaft (DFG, German Research Foundation) - grant number NI 2226/1-1 and Project-ID 499552394 – SFB 1597. The authors would like to thank Dr. Maria Petropoulou for providing comments on an early version of the manuscript.

\section{Data Availability Statement}

Code to reproduce the analysis and examples in Sections \ref{sec:database} and \ref{sec:examples} is available at \href{https://github.com/augustinewigle/poth-analysis}{https://github.com/augustinewigle/poth-analysis}. The R package \textbf{poth} is available at \href{https://github.com/augustinewigle/poth}{https://github.com/augustinewigle/poth}.

\bibliographystyle{apalike}
\bibliography{references}

\end{document}